\documentclass[pre,twocolumn,floatfix]{revtex4-2}
\usepackage{amsmath,amssymb}
\usepackage{amsthm}
\usepackage{graphicx}
\usepackage{xcolor}
\usepackage[normalem]{ulem}
\usepackage{etoolbox}

\makeatletter
\AtBeginDocument{%
  \patchcmd{\section}{\normalfont\small\bfseries}
    {\normalfont\small\bfseries\ifrevisionmarkup\color{revisionblue}\fi}{}
    {\PackageError{revision}{Section color patch failed}{Check the REVTeX section definition.}}%
}
\makeatother

\usepackage{xr}
\newif\ifrevisionmarkup
\revisionmarkuptrue
\definecolor{revisionblue}{RGB}{0,0,0}
\DeclareRobustCommand{\revadd}[1]{\ifrevisionmarkup{\color{black}#1}\else#1\fi}

\DeclareRobustCommand{\revcorechange}[2]{\ifrevisionmarkup{\color{black}#2}\else#2\fi}
\newenvironment{revisionblock}{\par\begingroup\ifrevisionmarkup\color{black}\fi}{\par\endgroup}

\begin{document}

\title{Periodic orbits underlying optical turbulence in the Lugiato–Lefever equation}

% \title{Periodic orbits underlying \revadd{weak spatiotemporal chaos} in the Lugiato–Lefever equation}

\author{Andrey Gelash$^{1}$}
\thanks{These authors contributed equally.}
\author{Savyaraj Deshmukh$^{2}$}
\thanks{These authors contributed equally.}
\author{Andrey Shusharin$^{1}$}
\author{Tobias M. Schneider$^{2}$}
\email[Corresponding author: ]{tobias.schneider@epfl.ch}
\author{Tobias J. Kippenberg$^{1}$}

\affiliation{$^{1}$Institute of Physics, Swiss Federal Institute of Technology Lausanne (EPFL), Lausanne, Switzerland}
\affiliation{$^{2}$Emergent Complexity in Physical Systems Laboratory, Swiss Federal Institute of Technology Lausanne (EPFL), Lausanne, Switzerland}

\begin{abstract} 
Optical turbulence in driven nonlinear resonators is a form of spatiotemporal chaos that remains poorly understood and is largely characterized through simulations and experiments but lacks a dynamical-systems description. Here we show that in the regime of weak spatiotemporal chaos, the dynamics is structured by families of exact, fully nonlinear, time-periodic solutions of the Lugiato–Lefever equation for driven-dissipative Kerr resonators. These unstable periodic orbits, which we term Oscillating Turing Rolls (OTR), are embedded within the spatiotemporal chaos of optical turbulence. Parametric continuation using Newton–Krylov methods reveals their existence range, stability, and bifurcation-theoretic origin. In symmetry-constrained domains, OTR solutions are intermittently shadowed by chaotic trajectories; in extended domains, they appear as localized motifs of spatiotemporal chaos. Identifying these nonlinear, non-chaotic building blocks establishes a dynamical-systems-based description of optical turbulence in nonlinear optical resonators.
\end{abstract}
\maketitle

\section{Introduction}
Many advances in photonics are rooted in the pattern-forming nature of nonlinear optical resonators. The Lugiato–Lefever equation (LLE) distills this physics and describes pattern formation in damped-driven optical Kerr cavities with applications ranging from fiber to chip-integrated optical resonators \cite{lugiato1987spatial,matsko2011mode,grelu2015nonlinear,Gorodetsky2018dissipative}.  
For the case of anomalous group velocity dispersion, the dimensionless LLE for the intracavity field $\psi(\theta,\tau)$ reads, 
\begin{equation}
    \label{eq:LLE}
    i \frac{\partial \psi}{\partial \tau} + \frac{1}{2} \frac{\partial^2 \psi}{\partial \theta^2} + |\psi|^2\psi = (-i+\zeta_0)\psi +if\,,
\end{equation}
where $\theta$ and $\tau$ are the intracavity spatial coordinate and time. 
The two control parameters are the pump-cavity detuning $\zeta_0$ and the pump laser strength $f$. 
The LLE constitutes a damped-driven nonlinear Schr\"odinger equation (NLSE) \cite{barashenkov1996existence}.
Perturbation theory thus allows us to transfer analytical solutions of the integrable NLSE \cite{zakharov1972exact} constructed via inverse scattering transforms to the small damping and forcing regime of the LLE \cite{kaup1978solitons,nozaki1986low,wabnitz1996control,coppini2020fermi}. In general, however, no analytical description 
of the nonlinear dynamics including coherent optical states exists and equation (\ref{eq:LLE}) must be solved numerically.

\begin{figure}[!t]\centering
	\includegraphics[width=8.6cm]{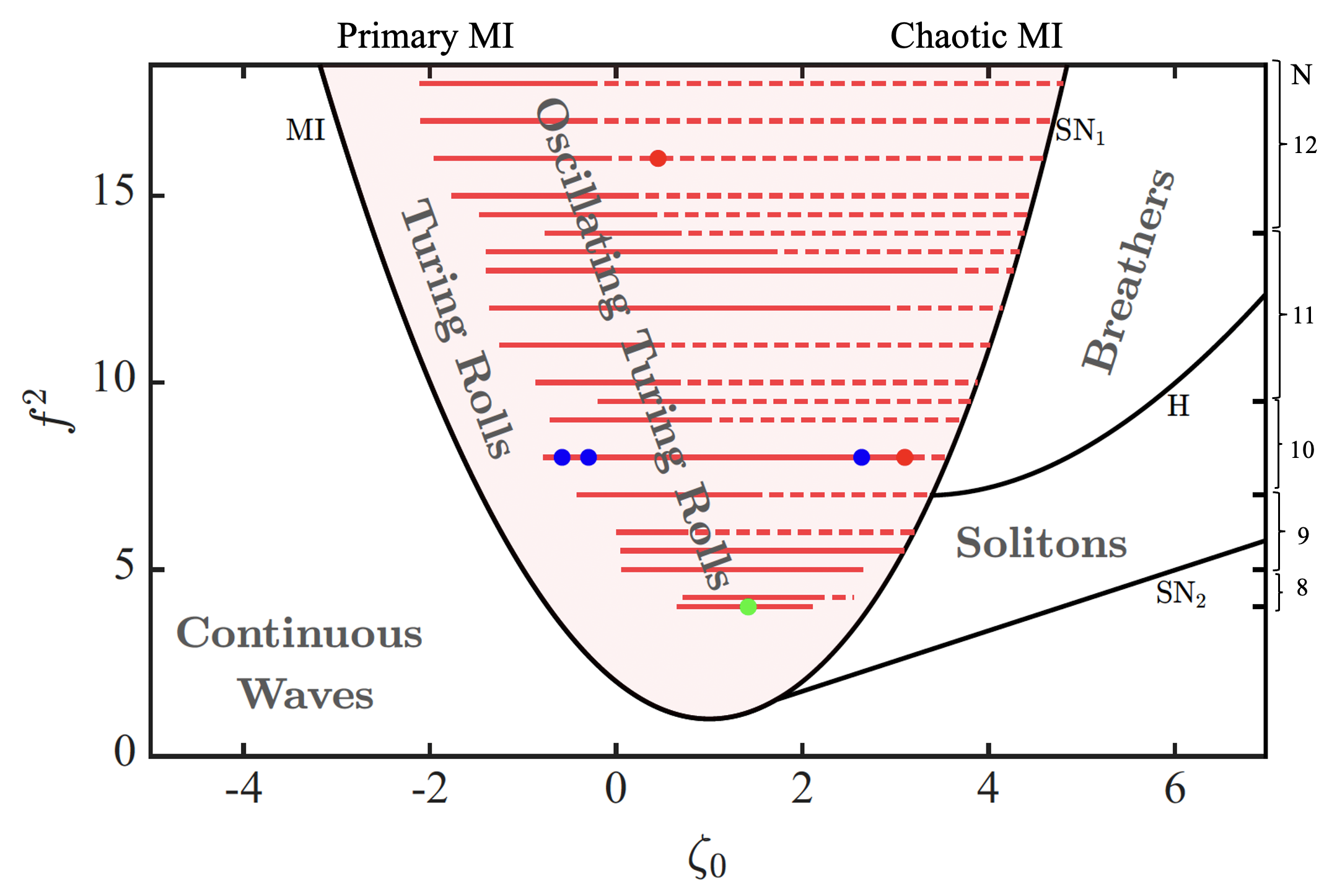}
    \caption{Existence and stability range of $\mathrm{OTR}_{\mathrm{I}}$ periodic orbit solutions \revadd{within discrete translational symmetry subspaces}, shown over the classical phase diagram of the LLE model (\ref{eq:LLE}). 
    The red solid (dashed) lines indicate stable (unstable) $\mathrm{OTR}_{\mathrm{I}}$ solution branches for 20 fixed values of $f^{2} \in [4,18]$, computed within the supercritical region of the $\mathrm{CW}$ bounded by the $\mathrm{MI}$ curve for $\zeta_0<2$ and the saddle-node bifurcation curve $\mathrm{SN_1}$ for $\zeta_0>2$. The number $N\in[8,12]$ indicates the specific discrete symmetry subspace $\theta \mapsto \theta + 16\pi/N$ used for each $f^2$. Dots ($\textcolor{blue}{\bullet}$, $\textcolor{red}{\bullet}$, $\textcolor{green}{\bullet}$) mark specific $\mathrm{OTR}_{\mathrm{I}}$ solutions discussed in the article.}
\label{fig:phase_diagram}
\end{figure}

Depending on the values of the control parameters ($\zeta_0, f$), several dynamical regimes have been characterized and are reliably observed in many optical resonator platforms \cite{herr2014temporal,leo2010temporal, Leo2013Dynamics,lucas2017breathing}. These include unpatterned spatially homogeneous continuous waves, stationary spatially periodic Turing rolls, stationary spatially localized solitons, and time-periodic spatially localized breathers \cite{godey2014stability}; \revadd{see Figure~\ref{fig:phase_diagram}}.
Stationary continuous waves (CW), Turing rolls (TR) and solitons have been described as exact equilibrium solutions of the LLE and their bifurcation structure has been thoroughly investigated~\cite{Parra-Rivas2018localized}. 
Time-periodic solutions corresponding to spatially localized breathers have also been computed and continued using a Newton method based on explicit Jacobian pseudoinversion, and their stability and bifurcations have been characterized through Floquet analysis~\cite{johansson2019stability}.

When the detuning parameter is increased, a modulational instability (MI)~\cite{parra2018bifurcation, qi2019dissipative} of the initially stable CW branch creates a stable TR branch. Further beyond the threshold, the TR branch itself loses stability and the system typically displays weak spatiotemporal chaos \cite{haelterman1992dissipative,haelterman1992low}, here referred to as \emph{`optical turbulence'}~\cite{coulibaly2019turbulence,coillet2019transition,gomila2003transition} in analogy with transitional hydrodynamic turbulence~\cite{eckhardt2007turbulence,Reetz2019,Graham2021}.
Various numerical and experimental studies have investigated this transition to chaos within the context of optical frequency comb formation. In the spectral domain, the transition from TR underlying primary combs is characterized by the broadening of MI gain lobes followed by secondary comb formation~\cite{herr2012universal,coillet2019transition,gomila2022role,coillet2014routes,panajotov2017spatiotemporal,cheng2025numerical, coulibaly2019turbulence} and eventually broadband optical turbulence.
Despite the growing interest in the fundamental understanding of optical turbulence as well as its increasing relevance for applications in frequency comb generation~\cite{coulibaly2019turbulence,lukashchuk2023chaotic,wang2025collision}, a theoretical description of the chaotic regime in terms of invariant solutions of the fully nonlinear LLE remains elusive.

Here we identify exact non-chaotic time-periodic solutions in the chaotic regime of the LLE and elucidate their bifurcation-theoretical origins. The identified unstable periodic orbits are shown to form the fundamental building blocks of both temporal and spatio-temporal chaos in driven optical resonators. The work is inspired by recent breakthroughs in the understanding of spatio-temporal patterns and transitional turbulence in shear flows, where the identification of dynamically unstable non-chaotic periodic orbit solutions embedded in the chaotic attractor of the 3D Navier-Stokes equations has established dynamical systems as a paradigm to study turbulence~\cite{hopf1948mathematical, Nagataf1990, Kerswell2005, Gibson2008, suri2017forecasting, Reetz2019, Graham2021}. The approach yields a picture of spatio-temporal chaos as a chaotic trajectory in the system's state space that transiently shadows periodic orbits.  

\section{Numerical methods}
Equilibria, periodic orbits, and relative periodic orbits are so-called \emph{invariant solutions} - due to their ability to capture patterns often also termed exact coherent states (ECS) - and satisfy the condition
\begin{equation}
    \label{eq:invariant}
     \revadd{\boldsymbol{\sigma}}\mathcal{F}^T (\boldsymbol{x}^*) - \boldsymbol{x}^* = 0,
\end{equation}
where $\boldsymbol{x}^*$ is a state vector along the orbit and $\mathcal{F}^T$ indicates the time evolution of a state $\boldsymbol{x}(\tau)$ to $\boldsymbol{x}(\tau+T) = \mathcal{F}^T (\boldsymbol{x}(\tau))$ according to the LLE (\ref{eq:LLE}). The operator $\sigma$ denotes a symmetry transformation, such as a discrete reflection or a spatial shift compensating the drift of a relative periodic orbit; for equilibria and ordinary periodic orbits, $\sigma$ is the identity. To numerically identify invariant solutions, track them in parameter space, characterize their dynamical stability and identify bifurcations, we transfer Newton-Krylov based continuation methods developed in the context of 3D fluid flows to the LLE system, as described in~\cite{Deshmukh2026} and summarized below.

\begin{figure*}[htbp]
\ifrevisionmarkup\color{revisionblue}\fi
\centering
	\includegraphics[width=17.5cm]{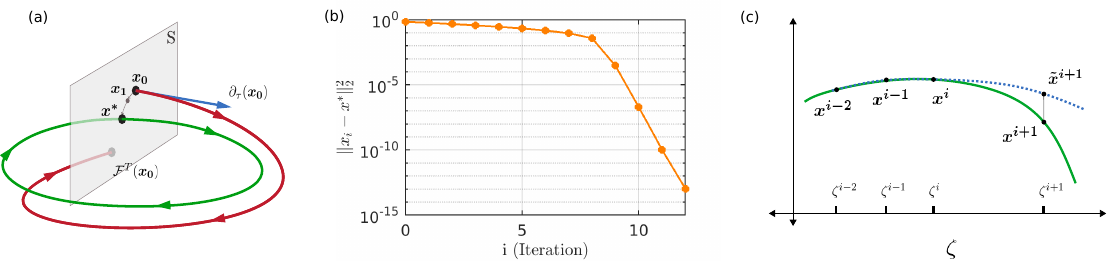}
	\caption{Newton--Krylov computation and parametric continuation of periodic orbits.
    (a) State-space schematic of Newton iterations toward a periodic orbit satisfying Eq.~\eqref{eq:invariant}. The initial guess $\boldsymbol{x}_0$ is iteratively updated on the Poincaré section $S$ until it converges to a fixed point of the return map Eq.~\eqref{eq:invariant}, $\boldsymbol{x}^*$, that lies on the periodic orbit (green).
    (b) Convergence of the Newton iterations, quantified by the squared state-space distance from the converged solution, $\|\boldsymbol{x}_i-\boldsymbol{x}^*\|^2_2$, at each Newton iteration $i$.
    (c) Schematic of natural parameter continuation of a solution branch (green) in a parameter $\zeta$. Quadratic extrapolation (blue) through three previously computed solutions predicts the guess $\tilde{\boldsymbol{x}}^{i+1}$ at $\zeta^{i+1}$, which is corrected by Newton iteration to obtain the solution $\boldsymbol{x}^{i+1}$.}
\label{fig:OTR_algorithms}
\end{figure*}

% Integrated from supplement: numerical methods and algorithm figure
\begin{revisionblock}
% \textit{Numerical methods.---} 
% Periodic orbit solutions, corresponding to fixed points $\boldsymbol{x}^*$ of the return map given by Eq.~\eqref{eq:invariant}, are computed using Jacobian-free Newton-Krylov algorithms, as described in~\cite{Deshmukh2026}. 
We discretize the spatial domain in Fourier space using 512 modes. (For the solutions in discrete symmetry subspaces discussed below, we use 128 modes.) 
The Newton-Krylov algorithm seeks a point $\boldsymbol{x}^*$ that returns to itself after evolution over time $T$ under the LLE flow map $\mathcal{F}^T$ and application of the symmetry operator $\sigma$ (Figure~\ref{fig:OTR_algorithms}(a)). Initial guesses for Newton’s method are obtained from long-time direct numerical simulations. If the system converges to a periodic attractor, this asymptotic state is used directly. In more complex regimes, transiently recurring periodic patterns are extracted from the time series and serve as effective initial conditions. For periodic orbits, the period $T$ is an additional unknown. For the relative periodic orbits discussed below, which return to their initial state up to a spatial translation, the spatial drift $\ell$ accumulated over one period is also an unknown. To obtain a system with equal number of constraints and unknowns, we require the Newton state correction to be orthogonal to the tangent directions $\partial_\tau\psi$ and, when needed, $\partial_\theta\psi$, fixing the temporal and spatial phases during the Newton iterations. These phase conditions augment the recurrence equation Eq.~\eqref{eq:invariant} to form a system $\boldsymbol{G}(\boldsymbol{y})=0$ for the combined unknowns $\boldsymbol{y}$, which Newton’s method updates via
\[
\boldsymbol{J}(\boldsymbol{y}_i) \, \delta \boldsymbol{y}_i = -\boldsymbol{G}(\boldsymbol{y}_i),
\]
where $\boldsymbol{J}$ is the Jacobian of the augmented residual $\boldsymbol{G}$. For computational efficiency, we avoid forming $\boldsymbol{J}$ explicitly and instead solve the linear system using Krylov subspace methods. These methods build low-dimensional subspaces spanned by successive applications of the Jacobian, evaluated through finite-difference approximations, thereby avoiding explicit construction of $\boldsymbol{J}$. We specifically employ the Generalized Minimal Residual (GMRES) algorithm~\cite{saad1986gmres}, which iteratively minimizes the residual over the Krylov subspace. To enhance convergence, we incorporate a trust-region strategy using the hookstep method~\cite{viswanath2007recurrent}, which restricts updates to remain within a bounded region where the linearization is accurate. The typical convergence behavior is shown in Figure~\ref{fig:OTR_algorithms}(b). We calculate the linear stability of the solutions of the return map in Eq.~\eqref{eq:invariant} using Arnoldi iteration, which provides the corresponding Floquet multipliers of the periodic orbits of Eq.~\eqref{eq:LLE}.
\end{revisionblock}

\begin{revisionblock}
To compute a solution branch under the variation of a parameter $\zeta$, we employ natural parameter continuation with quadratic extrapolation. Given three previously computed states $\boldsymbol{x}^{i-2}, \boldsymbol{x}^{i-1}, \boldsymbol{x}^{i}$ at parameter values $\zeta^{i-2}, \zeta^{i-1}, \zeta^{i}$, we construct a quadratic extrapolation $\boldsymbol{x}(\zeta)$ that fits the known data. The predicted state $\tilde{\boldsymbol{x}}^{i+1}$ at a new parameter value $\zeta^{i+1}$ is obtained by evaluating this extrapolation and then corrected using Newton iterations, including the period and phase conditions for periodic orbits, to solve
\[
\sigma\mathcal{F}^T(\boldsymbol{x}, \zeta) - \boldsymbol{x} = 0.
\]
This method provides an efficient way to trace smooth branches of solutions as a function of system parameters and thereby compute their bifurcation diagram; see Figure~\ref{fig:OTR_algorithms}(c).

When the solution branch bends back on itself (e.g., at saddle-node bifurcations), natural continuation is often inefficient and leads to many failures of Newton search attempts. In such cases, we use arclength continuation with an additional constraint:
\[
||\boldsymbol{x} - \boldsymbol{x}^i||^2 + (\zeta - \zeta^i)^2 = h^2,
\]
where $h$ is the arclength step. This constraint enforces the search for a new solution constrained within a sphere of radius $h$ around the extended state vector $\boldsymbol{\tilde{x}}^i= (\boldsymbol{x}^i, \zeta^i)$, allowing continuation past folds and turning points more efficiently.
\end{revisionblock}

\begin{figure}[!t]\centering
	\includegraphics[width=8.25cm]{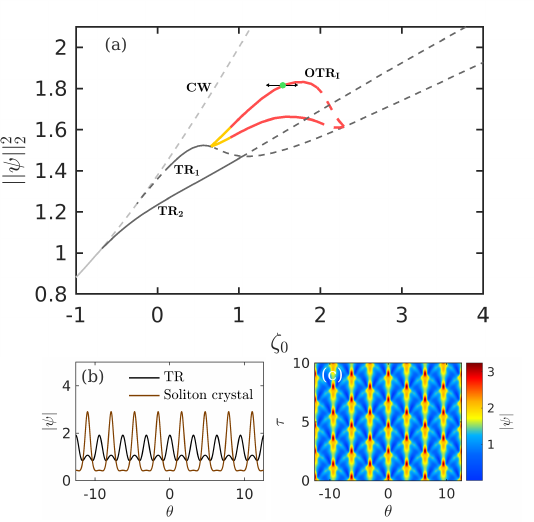}
	\caption{(a): Bifurcation diagram in the detuning parameter $\zeta_0$ for a fixed value of pump power $f^2=4$ and domain size $L=8\pi$. Solid (dashed) lines correspond to stable (unstable) solutions. It displays $\mathrm{CW}$ - continuous wave, $\mathrm{TR}_1$ and $\mathrm{TR}_2$ - Turing roll equilibrium solutions with $N=8$ and $N=9$ rolls respectively, $\mathrm{OTR}_{\mathrm{I}}$ - oscillating Turing roll periodic orbits (two curves represent the maxima and minima of the $L_2$-norm  $||\psi||_2^2=\frac{1}{L}\int_{-L/2}^{L/2}|\psi|^2d\theta$ during one period). Solid orange line corresponds to stable $\mathrm{OTR}_{\mathrm{I}}$, solid (dashed) red indicates stable (unstable) $\mathrm{OTR}_{\mathrm{I}}$ solutions restricted within a discrete symmetry subspace $\theta \mapsto \theta + 2L/8$. Green dot ($\textcolor{green}{\bullet}$) shows the starting point (also shown in Figure~\ref{fig:phase_diagram}) for the numerical continuation of the $\mathrm{OTR}_{\mathrm{I}}$ solution branch. (b): Spatial profile of solutions on the $\mathrm{TR}_1$ branch at $\zeta_0=0.498$ (MI region) corresponding to Turing rolls and $\zeta_0=3.706$ (Soliton region) corresponding to soliton crystals. (c): Spatiotemporal portrait of the $\mathrm{OTR}_{\mathrm{I}}$ solution at $\zeta_0=1.418$.}
\label{fig:bifurcation_diagram}
\end{figure}

\begin{figure*}[!t]\centering
	\includegraphics[width=17.5cm]{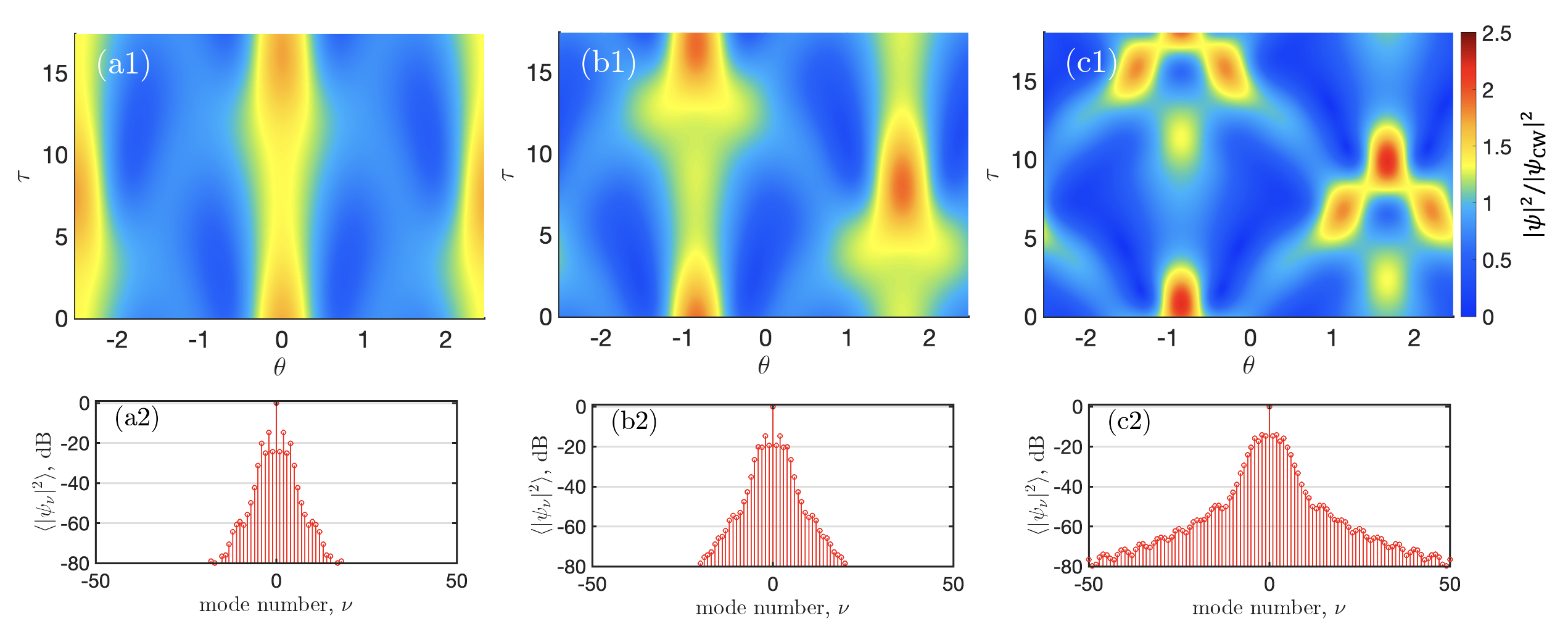}
	\caption{Examples of the $\mathrm{OTR}_{\mathrm{I}}$ solution for fixed pump power $f^2 = 8$ in a subspace of discrete shift symmetry $\theta \mapsto \theta + 2L/N$ with $N=10$ at detuning values $\zeta_0=-0.581$ (a), $\zeta_0=-0.297$ (b) and $\zeta_0=2.636$ (c) indicated by blue dots ($\textcolor{blue}{\bullet}$) in Figure~\ref{fig:phase_diagram}. Spatiotemporal portraits of normalized intensity $|\psi|^2/|\psi_{\mathrm{CW}}|^2$ (top) and time-average Fourier spectra (bottom) show that the anti-phase structure is preserved but the peak amplitude and width of the power spectrum increase for increasing detuning. 
    }
\label{fig:OTR_examples}
\end{figure*}

\section{Results}

In the low-dispersion limit Eq.(\ref{eq:LLE}) is considered on the real line $\theta \in \mathbb{R}$ reflecting the large scale separation between solution structures (given by the dispersion length scale) and the circumference of the optical resonator. 
To numerically represent the spatiotemporal dynamics, we choose a periodic domain $\theta \in [-L/2, L/2)$ with period $L$ sufficiently large.
While localized solitons exist within a stable CW background, chaotic modulations of rolls are expected when the CW background is unstable. We consequently study the supercritical region of the two-dimensional parameter space ($\zeta_0,f$), indicated in Figure~\ref{fig:phase_diagram}. For parameter values within this range, spatially periodic dynamically stable TR equilibrium solutions exist. Their wavelength tends to decrease with increasing pump power and consequently, the number of rolls $N$ on a periodic domain increases \cite{qi2019dissipative,skryabin2021threshold}. On increasing detuning for fixed $f$, TR solutions lose stability. 

We perform direct numerical simulations for $L=8\pi$ and vary $\zeta_0$ and $f$. 
Starting from low-amplitude noise, the dynamics either approaches the stable TR solution or - beyond their stability limit - we typically find temporal oscillations that either persist indefinitely or are transient and are followed by quasiperiodic or chaotic behavior. By extracting initial guesses from the oscillatory patterns and converging them using Newton's method, we identify several exact periodic orbit solutions capturing the observed spatiotemporal patterns. Three families of these $\mathrm{OTR}$ (Oscillating Turing Rolls) solutions are found: $\mathrm{OTR}_{\mathrm{I}}$ shows anti-phase oscillations of adjacent rolls, $\mathrm{OTR}_{\mathrm{II}}$ solutions are characterized by an $2\pi/3$ phase shift between adjacent rolls, and for $\mathrm{OTR}_{\mathrm{III}}$ adjacent rolls oscillate in phase. 

\begin{figure*}[htbp]
\ifrevisionmarkup\color{revisionblue}\fi
\centering
	\includegraphics[width=14.0cm]{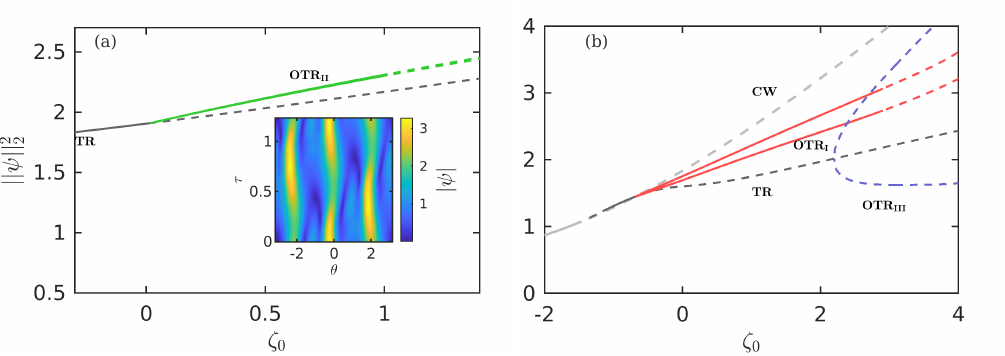}
	\caption{Bifurcation structure of the $\mathrm{OTR}_{\mathrm{II}}$ and $\mathrm{OTR}_{\mathrm{III}}$ solution families.
    (a) Continuation in detuning $\zeta_0$ at $f^2=16$ within the discrete symmetry subspace $\theta\mapsto\theta+3L/N$, with $L=8\pi$ and $N=12$. A stable $\mathrm{OTR}_{\mathrm{II}}$ relative periodic orbit branch bifurcates from the $\mathrm{TR}$ branch at $\zeta_0=0.017$. The inset shows the spatiotemporal profile of $\mathrm{OTR}_{\mathrm{II}}$ at $\zeta_0=0.45$.
    (b) Continuation at $f^2=8$ within $\theta\mapsto\theta+2L/N$, with $L=8\pi$ and $N=10$, showing the $\mathrm{OTR}_{\mathrm{I}}$ and $\mathrm{OTR}_{\mathrm{III}}$ solution branches. At $\zeta_0=2.2$, an unstable $\mathrm{OTR}_{\mathrm{III}}$ branch bifurcates from the unstable $\mathrm{TR}$ branch. Solid (dashed) lines denote stable (unstable) solutions within the specified symmetry subspaces. For periodic and relative periodic orbits, pairs of curves give the temporal maxima and minima of the spatially averaged intensity $\|\psi\|_2^2=\tfrac{1}{L}\int_{-L/2}^{L/2}|\psi|^2\,d\theta$. These extrema nearly coincide for $\mathrm{OTR}_{\mathrm{II}}$.}
\label{fig:bifurcation_family_2_3}
\end{figure*}

For fixed forcing $f^2=4$, numerical continuation of the $\mathrm{OTR}_{\mathrm{I}}$ solution branch in the detuning parameter reveals its emergence from an equilibrium $\mathrm{TR}$ solution via a Hopf bifurcation, as depicted in Figure~\ref{fig:bifurcation_diagram} (a). Specifically, $\mathrm{OTR}_{\mathrm{I}}$ bifurcates off $\mathrm{TR}_1$ with $N=8$ equidistant rolls, that itself bifurcates off CW at $\zeta_0=-0.25$. At onset $\mathrm{TR}_1$ is unstable, gains stability at $\zeta_0=0.13$, and the supercritical Hopf bifurcation at $\zeta_0=0.65$ creates a stable $\mathrm{OTR}_{\mathrm{I}}$ periodic orbit, which loses stability at $\zeta_0=0.80$.   
The amplitude of out-of-phase oscillations first increases as $\zeta_0$ is increased beyond the critical value of the Hopf bifurcation but then decreases again until the now unstable $\mathrm{OTR}_{\mathrm{I}}$ branch terminates on the (unstable) $\mathrm{TR}_1$ at $\zeta_0=2.29$. $\mathrm{OTR}_{\mathrm{I}}$ and $\mathrm{TR}_1$ coexist with $\mathrm{TR}_2$, a second equilibrium branch with $N=9$ rolls, that bifurcates off $\mathrm{CW}$ at $\zeta_0=-0.683$, is stable at onset and loses stability at $\zeta_0=1.03$. Both $\mathrm{TR}$ branches can be followed to higher values of detuning and, in the subcritical regime, where a stable $\mathrm{CW}$ exists, take the form of soliton crystals \cite{cole2017soliton}, as shown in Figure~\ref{fig:bifurcation_diagram}(b). 
$\mathrm{TR}_1$ is spatially periodic and invariant under shifts $\theta \mapsto \theta + L/N$ with $N = 8$. This shift symmetry is broken by the Hopf bifurcation generating $\mathrm{OTR}_{\mathrm{I}}$ and the resulting periodic orbit is invariant under shifts by $2L/N$, reflecting the anti-phase oscillations between adjacent rolls.  

\begin{figure*}[!t]\centering
	\includegraphics[width=17.5cm]{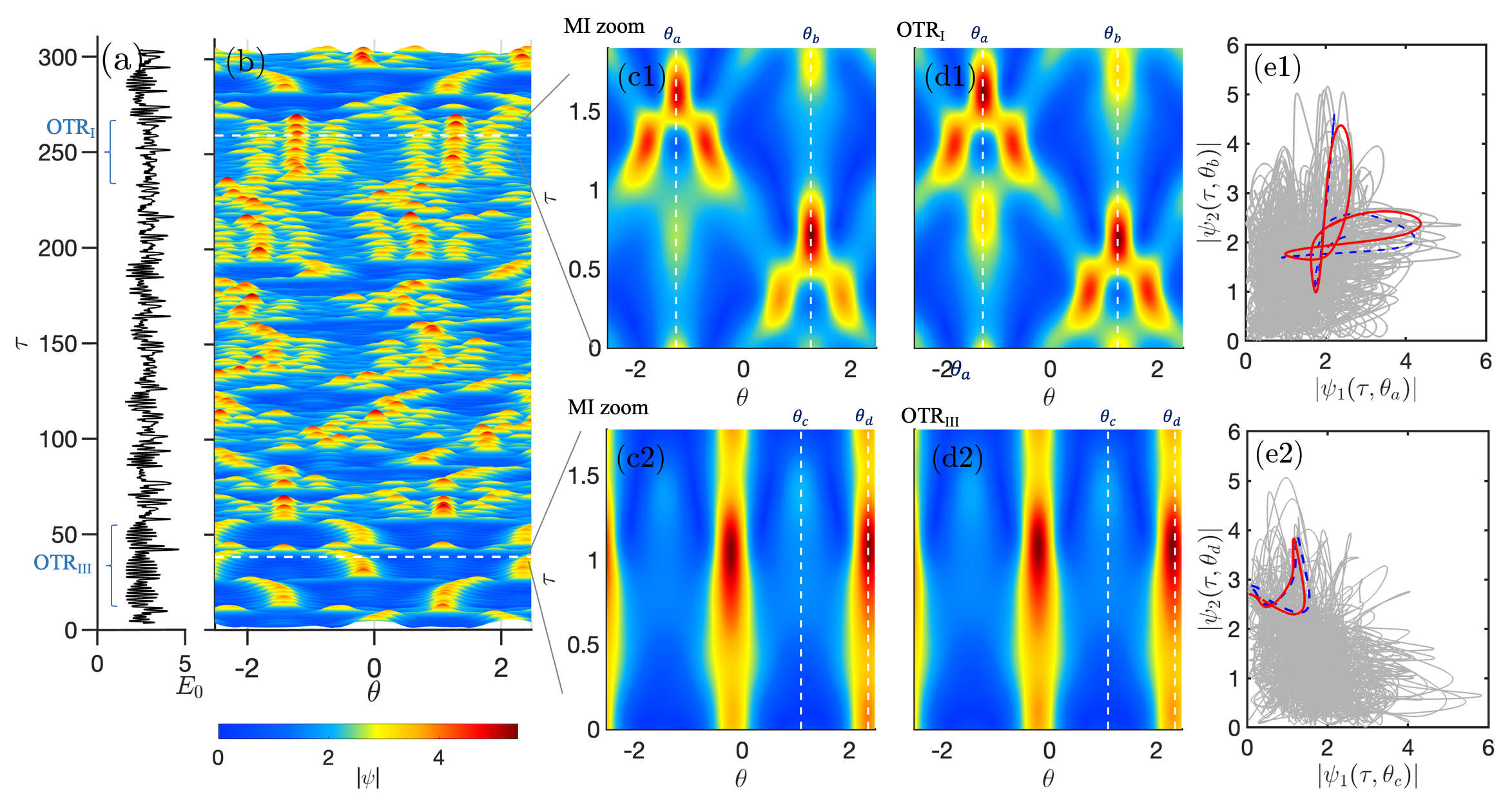}
	\caption{Unstable time-periodic $\mathrm{OTR}_{\mathrm{I}}$ and $\mathrm{OTR}_{\mathrm{III}}$ solutions are transiently visited by the temporally chaotic dynamics in the $\theta \mapsto \theta + 2L/10$ symmetric subspace for $f^2 = 8$ and $\zeta_0=3.1$. 
    (a) Evolution of total intracavity power $E_0$ featuring regular and irregular time behaviour. (b) Spatiotemporal evolution of the intracavity field indicating intermittent switching between $\mathrm{OTR}_{\mathrm{I}}$, $\mathrm{OTR}_{\mathrm{III}}$ and chaos. The spatio-temporal signal, magnified in (c1, c2) closely and quantitatively resembles the exact periodic orbits $\mathrm{OTR}_{\mathrm{I}}$ and $\mathrm{OTR}_{\mathrm{III}}$ in (d1, d2). Further evidence that  $\mathrm{OTR}$ periodic orbits are indeed \revcorechange{shaddowed}{shadowed} is given by two-dimensional phase space portraits (e1, e2), where the dynamics is projected on coordinates given by the values of the \revcorechange{wave intensities}{field amplitudes} $|\psi|$ at fixed $\theta$ (white dashed lines in panels (c) and (d)). Segments of the chaotic trajectory (grey), indicated by dashed blue lines, follow the OTR periodic orbits (red), illustrating transient shadowing and their dynamical relevance.
    }
\label{fig:OTR_switching}
\end{figure*}

To identify \revcorechange{$\mathrm{OTR}$}{$\mathrm{OTR}_{\mathrm{I}}$} solutions for other fixed values of $f^{2} \in [4,18]$, we consider specific discrete translational symmetry subspaces based on the number of rolls $N$, that are typically observed in simulated weak chaos within the $L=8\pi$ domain. The number of rolls tends to increase with forcing. For a fixed value of $f^2$ we conduct a DNS in the subspace of discrete shift symmetry $\theta \mapsto \theta + 2L/N$, i.e. in a periodic domain of length $2L/N$. For detunings sufficiently beyond the MI threshold, simulations initialized with noise yield temporally periodic rolls oscillating in anti-phase. From those initial guesses, we converge $\mathrm{OTR}_{\mathrm{I}}$ solutions and numerically continue them towards smaller detuning parameters. The $\mathrm{OTR}_{\mathrm{I}}$ branches bifurcate off the corresponding $\mathrm{TR}$ equilibrium branch either via a supercritical Hopf bifurcation or, alternatively, via a subcritical Hopf bifurcation followed by a saddle-node on the $\mathrm{OTR}_{\mathrm{I}}$ branch. Consequently, the lower boundary in detuning for which $\mathrm{OTR}_{\mathrm{I}}$ exists, is given by the Hopf point in the supercritical and the saddle-node in the subcritical case. 
The upper boundary of the existence range is likewise demarcated by the $\mathrm{OTR}_{\mathrm{I}}$ branch terminating on the $\mathrm{TR}$ branch either directly or in a subcritical Hopf bifurcation preceded by a saddle-node bifurcation. For some values of $f^2$, the $\mathrm{OTR}_{\mathrm{I}}$ persists beyond the supercritical region into the breather regime where it takes the form of a \emph{breather crystal} -- an oscillating analogue of the soliton crystal. 
We report both the existence range in detuning as well as the wavelength of all identified $\mathrm{OTR}_{\mathrm{I}}$ periodic orbit solutions for varying $f^2$ in Figure~\ref{fig:phase_diagram}, along with their dynamical stability features computed using a Floquet analysis within the corresponding discrete symmetry subspaces of the solutions. All $\mathrm{OTR}_{\mathrm{I}}$ solutions show the characteristic antiphase oscillations of rolls with large detunings promoting large-amplitude peaks, as exemplified in Figure~\ref{fig:OTR_examples}.

% Integrated from supplement: OTR II/III branches and bifurcation figure
\begin{revisionblock}
A bifurcation analysis of the $\mathrm{OTR}_{\mathrm{II}}$ and $\mathrm{OTR}_{\mathrm{III}}$ solutions in $\zeta_0$ at selected pump powers reveals their origin from equilibrium $\mathrm{TR}$ solution branches. 
For $f^2=16$ and within a discrete symmetry subspace $\theta \mapsto \theta + 3L/N$ with $N=12$ (Figure~\ref{fig:bifurcation_family_2_3} (a)), an equilibrium $\mathrm{TR}$ solution loses stability at $\zeta_0=0.017$, giving rise to a stable $\mathrm{OTR}_{\mathrm{II}}$ relative periodic orbit branch (periodic orbit in a moving frame of reference~\cite{Avila2013}). 
As $\zeta_0$ is increased further, the $\mathrm{OTR}_{\mathrm{II}}$ solution loses stability at $\zeta_0=1.04$. 
For $f^2=8$ and within a discrete symmetry subspace $\theta \mapsto \theta + 2L/N$ with $N=10$, an unstable $\mathrm{OTR}_{\mathrm{III}}$ branch bifurcates from the unstable $\mathrm{TR}$ branch at $\zeta_0=2.2$ (Figure~\ref{fig:bifurcation_family_2_3} (b)).
\end{revisionblock}

\begin{figure*}[!t]
\ifrevisionmarkup\color{revisionblue}\fi
\centering
	\includegraphics[width=17.0cm]{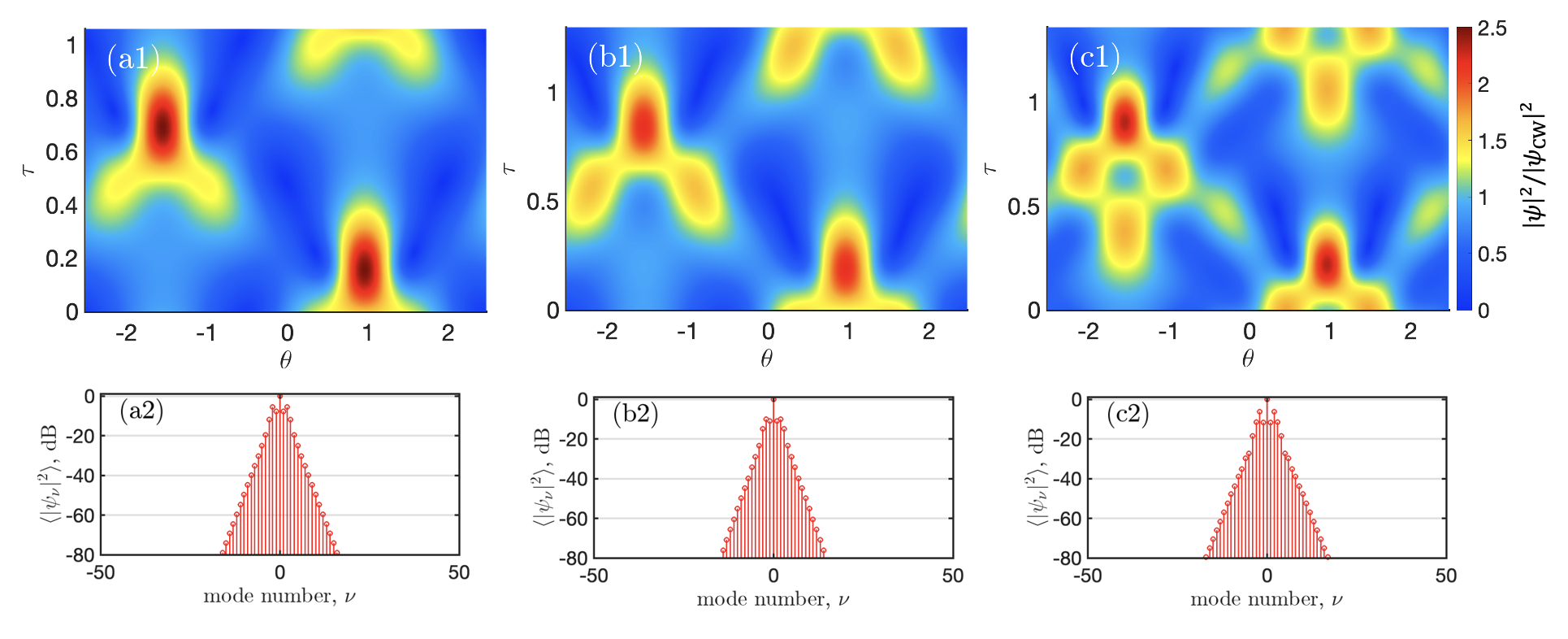}
	\caption{Dependence of $\mathrm{OTR}_{\mathrm{I}}$ solutions on spatial period at fixed $f^2=8$ and $\zeta_0=0.4$.
    Columns (a--c) correspond to periodic domain lengths $0.95L_0$, $L_0$, and $1.07L_0$, respectively, where the reference length is $L_0=2L/N$ with $L=8\pi$ and $N=10$.
    Panels (a1--c1) show the spatiotemporal profile of the normalized intensity $|\psi|^2/|\psi_{\mathrm{CW}}|^2$; panels (a2--c2) show the corresponding time-averaged Fourier power spectra in decibels as functions of the mode index $\nu$.}
\label{fig:OTR_examples_S1}
\end{figure*}

% \section{Regime of weak spatiotemporal chaos}

To study the dynamical relevance of the identified $\mathrm{OTR}$ periodic orbit solutions and specifically their significance for the onset of chaos, we consider the dynamics within the translational symmetry subspace associated with the $\mathrm{OTR}_{\mathrm{I}}$, thereby suppressing large-scale spatial variations. 
For forcing $f^{2}=8$ at a detuning just beyond the loss of stability of $\mathrm{OTR}_{\mathrm{I}}$, indicated in Figure~\ref{fig:phase_diagram}, all known steady and time-periodic solutions—including $\mathrm{TR}$, $\mathrm{OTR}_{\mathrm{I}}$, \revcorechange{as well as an additional $\mathrm{OTR}_{\mathrm{III}}$}{and $\mathrm{OTR}_{\mathrm{III}}$} are unstable, and DNS simulations of the system in $\theta \mapsto \theta + 2L/10$ symmetric subspace exhibit chaos (Figure~\ref{fig:OTR_switching}(b)). Despite the chaotic evolution, the dynamics intermittently displays oscillatory patterns (Figure~\ref{fig:OTR_switching}(c1, c2)) that closely resemble the $\mathrm{OTR}_{\mathrm{I}}$ and $\mathrm{OTR}_{\mathrm{III}}$ solutions (Figure~\ref{fig:OTR_switching}(d1,d2)). This suggests close visits of the unstable $\mathrm{OTR}$ periodic orbit solutions. Further evidence that the chaotic trajectory transiently visits and shadows both $\mathrm{OTR}_{\mathrm{I}}$ and $\mathrm{OTR}_{\mathrm{III}}$ solutions is provided by 2D state-space projections, shown in Figure~\ref{fig:OTR_switching}(e1,e2). 
These observations demonstrate that the OTR solutions are not merely isolated invariant states but are dynamically relevant structures that are repeatedly visited by chaotic trajectories and thus contribute to the organization of the turbulent dynamics.
Signatures of the chaotic trajectory closely visiting unstable $\mathrm{OTR}$ solutions are also present in the total intracavity power $E_0(\tau) = \int_0^L |\psi(\theta, \tau)|^2 d\theta$ (Figure~\ref{fig:OTR_switching}(a)), which shows periodic episodes within the otherwise chaotic temporal signal. This irregular switching between different time-periodic patterns has previously also been observed in a 4-dimensional Galerkin approximation of the LLE~\cite{haelterman1992low}.

\begin{figure*}[!t]\centering
	\includegraphics[width=17.5cm]{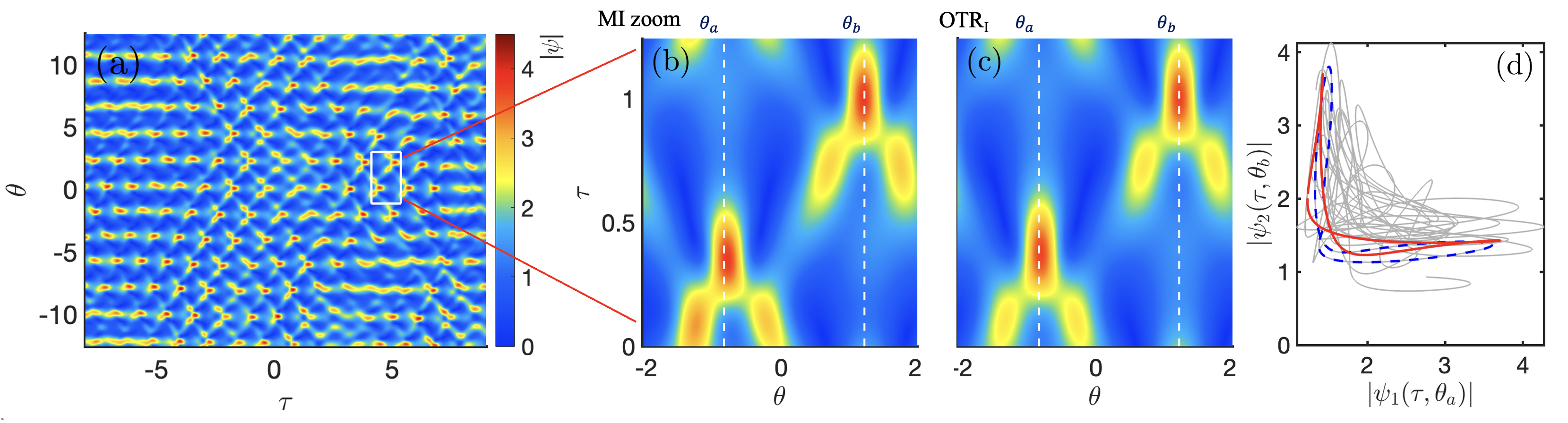}
	\caption{ 
    Oscillating roll patterns captured by $\mathrm{OTR}$ solutions spontaneously emerge within the spatio-temporal chaos in the full spatial domain of size $L=8\pi$ at $f^2=16$ and $\zeta_0=0.45$ (this point is indicated by red dot ($\textcolor{red}{\bullet}$) in Figure~\ref{fig:phase_diagram}) shown in (a). An example of the pattern, coexisting with other structures in space is magnified in (b). This pattern is very well described by the $\mathrm{OTR}_{\mathrm{I}}$ periodic orbit (c). In the full domain, the spatial period is not constrained, and thus the $\mathrm{OTR}_{\mathrm{I}}$ solution is shown for the measured spatial periodicity $\theta \mapsto \theta + 4.14$ of the pattern, obtained via continuation in the spatial period of the computational domain. (d): phase portrait as in Figure~\ref{fig:OTR_switching} panels (e1, e2) further suggests that the dynamics (blue dashed) follows the $\mathrm{OTR}$ periodic orbit (red).}
\label{fig:OTR_chaos}
\end{figure*}

$\mathrm{OTR}$ periodic orbit solutions are transiently visited building blocks of temporal chaos within a constrained symmetry subspace. To assess their significance also for spatiotemporally chaotic patterns in an extended domain, we consider a domain of size $L=8\pi$ at parameter values where all known simple invariant solutions, including $\mathrm{CW}$, $\mathrm{TR}$ and the $\mathrm{OTR}_{\mathrm{I}}$ have lost stability. Space-time plots of the dynamics, such as those shown in Figure~\ref{fig:OTR_chaos}(a) for parameter values $f^2=16$ and $\zeta_0=0.45$, \revcorechange{indicates}{indicate} distinct approximately time-periodic patterns in localized regions of the domain\revcorechange{, which closely resemble the $\mathrm{OTR}_{\mathrm{I}}$ solution for the same parameter values.}{. While these patterns closely resemble the $\mathrm{OTR}_{\mathrm{I}}$ solution at the same parameter values, their local spatial period is not prescribed in the extended domain and differs slightly from the spatial period $L_0=2L/N$ of the computed solution.} \revadd{We perform numerical continuation of the $\mathrm{OTR}_{\mathrm{I}}$ solutions in $L_0$ at fixed detuning and forcing, revealing that their spatiotemporal profiles and characteristic maximum field amplitudes vary considerably even with small changes in $L_0$ (Figure~\ref{fig:OTR_examples_S1}).} 

\revcorechange{}{For $f^2=16$ and $\zeta_0=0.45$, the $\mathrm{OTR}_{\mathrm{I}}$ orbit obtained at the measured local spatial period $L_0=4.14$ closely resembles the oscillatory pattern in the extended-domain simulation (Figure~\ref{fig:OTR_chaos}(b,c)).} Consequently, spatially periodic $\mathrm{OTR}$ solutions capture `\revcorechange{motives}{motifs}' of the spatiotemporal dynamics and describe observed patterns locally, yet alone they cannot describe the coexistence of different patterns emerging simultaneously in different parts of the spatial domain. Following arguments in other spatially extended pattern forming systems~\cite{Cross2009, Subramanian2021} we may however consider coexisting spatially periodic $\mathrm{OTR}$ solutions as `building blocks' of the dynamics. Together with temporally evolving fronts -- or defects -- between different regular patterns, they support the spatio-temporal dynamics.

\section{Conclusions}

We have identified unstable time-periodic solutions of the LLE in the modulationally unstable parameter regime. These unstable $\mathrm{OTR}$ periodic orbit solutions capture Oscillating Turing Roll patterns and constitute building blocks of chaos; they are transiently visited during temporally chaotic dynamics in discrete translational symmetry subspaces and capture \revcorechange{motives}{motifs} of optical turbulence in extended domains. 
While we identified $\mathrm{OTR}$ solutions in the low-dispersion limit of the LLE, they should also be observable in \revcorechange{microresonators experiments}{microresonator experiments} with large dispersion because the LLE restricted to a discrete translational-symmetry subspace $\theta \mapsto \theta + L_0$ of the infinite domain in the low-dispersion limit maps exactly onto the LLE on a finite periodic domain of size $L_0$ with a normalized dispersion coefficient $d_2 = \left(2\pi/L_0\right)^2$. In the large-dispersion regime, temporal high-resolution measurements of the intracavity power \cite{lucas2017breathing} should suffice to directly observe $\mathrm{OTR}$s. Recently developed techniques for measuring the spatially resolved \revcorechange{intercavity field}{intracavity field} \cite{wang2025collision} offer a route to observing $\mathrm{OTR}$ solutions in the low-dispersion limit as well.

\revadd{Although we have characterized three $\mathrm{OTR}$ families, we expect a much larger set of unstable periodic orbits to underlie the chaotic dynamics. Identifying these orbits could enable a description of spatiotemporal chaos in extended domains as a ``spatiotemporal tapestry'' of coherent subdomains locally shadowing different periodic orbits, interspersed with chaotic subdomains and connected by evolving interfaces~\cite{cvitanovic2026chaotic,zhigunov2026exact}. Such a description could shed light on the transition from spatiotemporal intermittency to fully developed amplitude turbulence observed at stronger driving~\cite{coulibaly2019turbulence}. The length and time scales of coherent subdomains could also be investigated by comparing Bloch--Floquet instability wavelengths and growth times of the $\mathrm{OTR}$ solutions~\cite{perinet2017eckhaus,gomila2022role,parra2018bifurcation} with coherent-patch size and lifetime statistics from an ensemble of direct numerical simulations.} While we uncovered the existence and bifurcation structure of unstable periodic orbit solutions underlying the chaotic dynamics within a single driven Kerr resonator, the general approach naturally extends to more complex optical resonators including dispersion-engineered and coupled systems \cite{tikan2021emergent,anderson2023dissipative,lucas2023tailoring,tusnin2020nonlinear,englebert2023bloch, villois2023walk}, as well as spatially modulated pumping schemes where periodic solutions of the LLE have recently been observed \cite{sun2025dissipative}.

\begin{acknowledgments}
This work was supported by funding from the Swiss National Science Foundation under grant agreement No. 216493 (HEROIC) and the European Research Council (ERC) under the European Union’s Horizon 2020 research and innovation programme (Grant No. 865677). 
\end{acknowledgments}

%%%%%%%%%%%%%%%%%
%%%%% BIBLIOGRAPHY

\end{document}